\documentclass[prl,reprint,nofootinbib,notitlepage,floatfix,superscriptaddress]{revtex4-1}
\usepackage[pdftex]{graphicx}

\usepackage{amsmath,amssymb,amstext}
\usepackage{mathrsfs}

\usepackage{multirow}

\usepackage{pdfsync}

\usepackage{comment}
\usepackage{bbold}
\usepackage{dsfont}
\usepackage{color}
\usepackage{hyperref}

\newcommand{\ket}[1]{\left|#1\right>} 

\begin{document}
  \title{Direct characterization of ultrafast energy-time entangled photon pairs}

  \author{Jean-Philippe W. MacLean}
  \email{jpmaclean@uwaterloo.ca}
  \affiliation{Institute for Quantum Computing, University of Waterloo, Waterloo,
  Ontario, Canada, N2L 3G1} 
  \affiliation{Department of Physics \& Astronomy, University of Waterloo,
  Waterloo, Ontario, Canada, N2L 3G1}
  \author{John M. Donohue}
  \affiliation{Institute for Quantum Computing, University of Waterloo, Waterloo,
  Ontario, Canada, N2L 3G1} 
  \affiliation{Department of Physics \& Astronomy, University of Waterloo,
  Waterloo, Ontario, Canada, N2L 3G1}
  \affiliation{Integrated Quantum Optics, Applied Physics, University of
  Paderborn, 33098 Paderborn, Germany} 
  \author{Kevin J. Resch}
  \affiliation{Institute for Quantum Computing, University of Waterloo, Waterloo,
  Ontario, Canada, N2L 3G1} 
  \affiliation{Department of Physics \& Astronomy, University of Waterloo,
  Waterloo, Ontario, Canada, N2L 3G1}

  \date{\today} 
  
  \begin{abstract} 
    Energy-time entangled photons are critical in many quantum optical
    phenomena and have emerged as important elements in quantum information
    protocols.  Entanglement in this degree of freedom often manifests itself
    on ultrafast timescales making it very difficult to detect, whether one
    employs direct or interferometric techniques, as photon-counting detectors
    have insufficient time resolution.  Here, we implement ultrafast photon
    counters based on nonlinear interactions and strong femtosecond laser
    pulses to probe energy-time entanglement in this important regime.  Using
    this technique and single-photon spectrometers, we characterize all the
    spectral and temporal correlations of two entangled photons with
    femtosecond resolution.  This enables the witnessing of energy-time
    entanglement using uncertainty relations and the direct observation of
    nonlocal dispersion cancellation on ultrafast timescales.  These techniques
    are essential to understand and control the energy-time degree of freedom
    of light for ultrafast quantum optics. 
  \end{abstract}
   
  \maketitle
 
 The energy-time degree of freedom of non-classical light is of great interest
 for quantum information as it supports various encodings, including frequency
 bins~\cite{ramelow_discrete_2009}, time bins~\cite{marcikic_time-bin_2002},
 and broadband temporal modes~\cite{brecht_photon_2015}, and is intrinsically
 robust for propagation through long-distance fibre
 links~\cite{zhang_distribution_2008}.  Applications which harness quantum
 correlations in this degree of freedom, referred to as energy-time
 entanglement~\cite{franson_bell_1989}, include dispersion
 cancellation~\cite{franson_nonlocal_1992,steinberg_dispersion_1992},
 high-dimensional quantum key distribution~\cite{nunn_large-alphabet_2013,
 lukens_orthogonal_2014}, and quantum-enhanced clock
 synchronization~\cite{giovannetti_quantum-enhanced_2001}.  In ultrafast optics
 and attosecond physics, the ability to measure both frequency and temporal
 features has led to important innovations in electric field reconstruction
 techniques~\cite{trebino_measuring_1997, walmsley_characterization_2009} and
 pulse characterization on very short timescales, enabling advances in
 spectroscopy~\cite{zewail_femtochemistry:_2000}, laser
 physics~\cite{bahk_generation_2004}, nonlinear
 optics~\cite{chang_generation_1997}, and imaging~\cite{zipfel_nonlinear_2003}.
 In order to characterize and control energy-time entangled photons and advance
 biphoton pulse shaping, similar measurement capabilities are essential in the
 quantum regime.

 Experimental signatures of entanglement can arise in correlation measurements
 of complementary variables~\cite{mancini_entangling_2002}, or through nonlocal
 quantum effects~\cite{franson_bell_1989,franson_nonlocal_1992}.  With the
 energy-time degree of freedom, one complementary set consists of measuring the
 intensity correlations as a function of the photon frequencies and as a
 function of their time of arrival.  These have been individually realized for
 different photonic systems with measurements in
 frequency~\cite{avenhaus_fiber-assisted_2009, schwarz_experimental_2014} or in
 time~\cite{kuzucu_joint_2008, shalm_three-photon_2013, cho_engineering_2014}.
 Certifying the presence of entanglement with direct measurements requires both
 spectral and temporal correlations, since acquiring only one remains
 insufficient to uniquely specify the other due to the ambiguity of the
 spectral phase.  Depending on the platform, this can be challenging.
 Narrowband photons from atomic systems can be readily measured in time but are
 difficult to spectrally resolve~\cite{cho_engineering_2014}. THz-bandwidth
 photons produced in spontaneous parametric downconversion (SPDC) are often
 characterized spectrally, but they can have features on femtosecond timescales
 below current detector resolution~\cite{eisaman_invited_2011}. 

 Other techniques can be employed to infer the presence of energy-time
 entanglement. High-order interference effects with Franson interferometers
 have been used to illustrate entanglement between
 two~\cite{kwiat_high-visibility_1993} and three
 photons~\cite{agne_observation_2017}. Nonlocal dispersion
 cancellation~\cite{franson_nonlocal_1992}, whereby the temporal spread in
 coincidences remains unchanged when equal and opposite dispersion is applied
 to each photon, can also be used to witness
 entanglement~\cite{wasak_entanglement-based_2010,
 jaramillo-villegas_persistent_2017}.  For either method to be effective,  the
 detector resolution must be shorter than the timescales of the
 correlations. 
 Strong energy-time entanglement can nonetheless exist when the timescales of
 the correlations are shorter.
 Certain observations have pointed to
 nonlocal dispersion cancellation in this regime, but they either
 required introducing a very large amount of dispersion such that temporal
 resolution could be achieved with standard
 detectors~\cite{baek_nonlocal_2009}, or used sum-frequency generation (SFG)
 between the photons pairs~\cite{odonnell_observations_2011}, which, unlike
 measurements with fast and independent detectors, has a close classical
 analogue~\cite{prevedel_classical_2011}.  Directly measuring ultrafast quantum
 effects requires new methods to control and analyze single photons in the time
 domain.  

   \begin{figure*}[t!]
     \centering
     \includegraphics[scale=1.0]{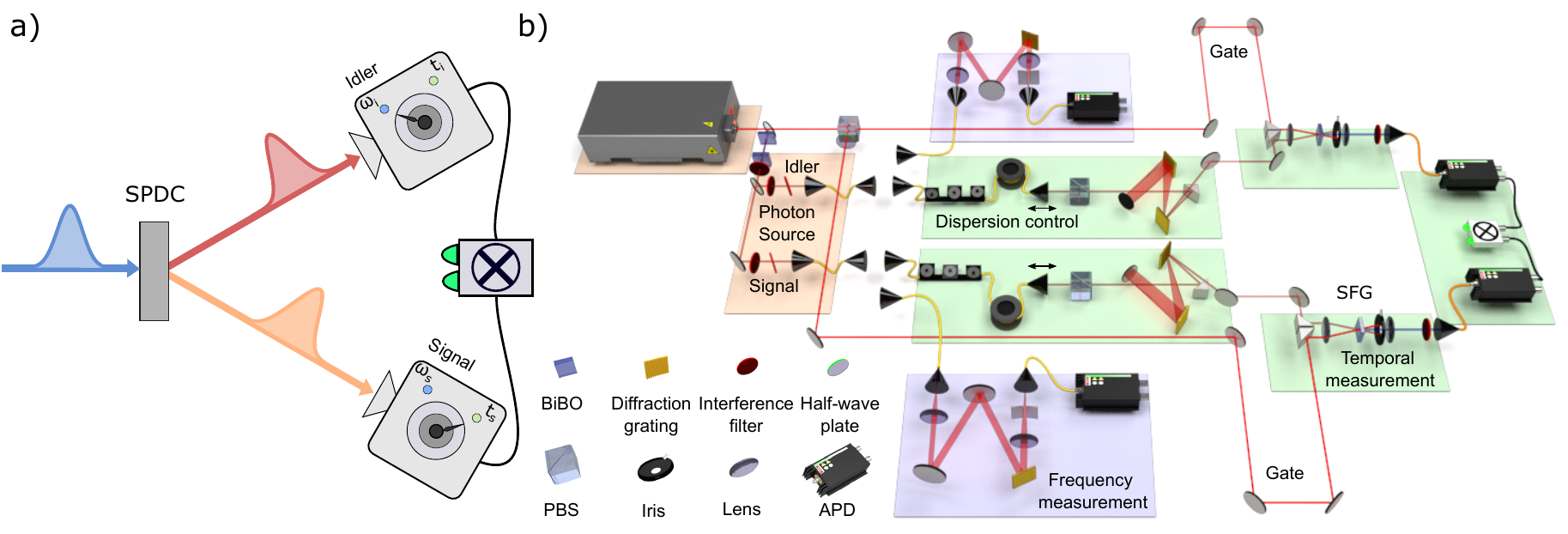}
     \caption{\footnotesize{ \textbf{Experimental setup.}
     (a) Frequency-entangled photons are created through spontaneous parametric
     downconversion of an ultrafast pulse from a frequency doubled Ti:sapphire
     laser. Measurements of either the frequency or the time of arrival of each
     photon can be performed in coincidence. (b) Spectral measurements are made
     with dual single-photon monochromators. Temporal measurements are
     performed using optically gated single-photon detection. The gating is
     implemented via noncollinear sum-frequency generation between a strong
     gate pulse from the Ti:Sapph laser and the signal or idler. The dispersion
     of the signal and idler photons is controlled with a combination of single
     mode fibres and grating compressors before the upconversion. The
     upconverted signal is filtered with bandpass filters which remove the
     background second harmonic generation from the gate pulse. Temporal and
     frequency measurements are performed in coincidence to observe the
     spectral and temporal features of the photons.}} 
     \label{fig:setup} 
  \end{figure*}

 In nonlinear optics and laser physics, optical gating is widely used to
 overcome limitations with detectors which are too slow to observe features on
 subpicosecond timescales.  The gating is achieved by combining the signal with
 a short gate pulse in a nonlinear medium and measuring the upconversion signal
 on the detector. With fast gates and slow detectors, an effective fast
 detector can be engineered to temporally resolve single
 photons~\cite{kuzucu_time-resolved_2008,allgaier_fast_2017} and photon
 pairs~\cite{kuzucu_joint_2008}.   In this work, we develop fast optical gating
 to achieve subpicosecond timing resolution for spatially separated pairs of
 single photons.  We use this technique in conjunction with single-photon
 spectrometers to  explicitly measure both the spectral and temporal
 correlations of broadband photons, as well as the cross-correlations between
 the frequency of one photon and time of arrival of the other.  Furthermore, by
 controlling the dispersion of each photon, our high-resolution joint temporal
 measurements make it possible to directly observe nonlocal dispersion
 cancellation on femtosecond timescales. 

 Through spectral and temporal measurements, energy-time entanglement can be
 witnessed by violating uncertainty relations~\cite{howell_realization_2004,
 edgar_imaging_2012}.  Two separable photons or classical pulses must satisfy
 the following
 inequality~\cite{mancini_entangling_2002,shalm_three-photon_2013},
  \begin{align}
    \Delta(\omega_s+\omega_i)\Delta(t_s-t_i)\geq1,
    \label{eq:TBP-inequality}
  \end{align}
 where each photon, labelled signal and idler, is described by its frequency
 $\omega$ and its time of arrival $t$, and $\Delta$ represents the standard
 deviation in the joint spectrum or joint temporal intensity.  In other words,
 there is a nontrivial limit to the strength of the product of correlations
 between the sum of the frequencies and the difference in time of arrival if
 the photons are separable. However, this is not the case for energy-time
 entangled photons where the right side of Eq.~\ref{eq:TBP-inequality} can approach
 zero. Thus, the uncertainty relation is an entanglement witness.
  
 Two-photon states produced via SPDC are often energy-time entangled.  In
 downconversion, energy conservation tends to lead to entangled states with
 frequency anti-correlations, although engineering SPDC sources have been
 explored to produce photon pairs with uncorrelated~\cite{chen_efficient_2017}
 or even positively correlated
 frequencies~\cite{grice_spectral_1997,grice_eliminating_2001,eckstein_highly_2011,
 harder_optimized_2013,donohue_spectrally_2016}.  For a pure state with no
 spectral phase, strong frequency correlations imply strong correlations in the
 time of arrival of the photons. Under these conditions,
 Eq.~\ref{eq:TBP-inequality} can be violated provided one has sufficient
 resolution in the measurements. 

\begin{figure}[t!]
     \centering
     \includegraphics[scale=1.0]{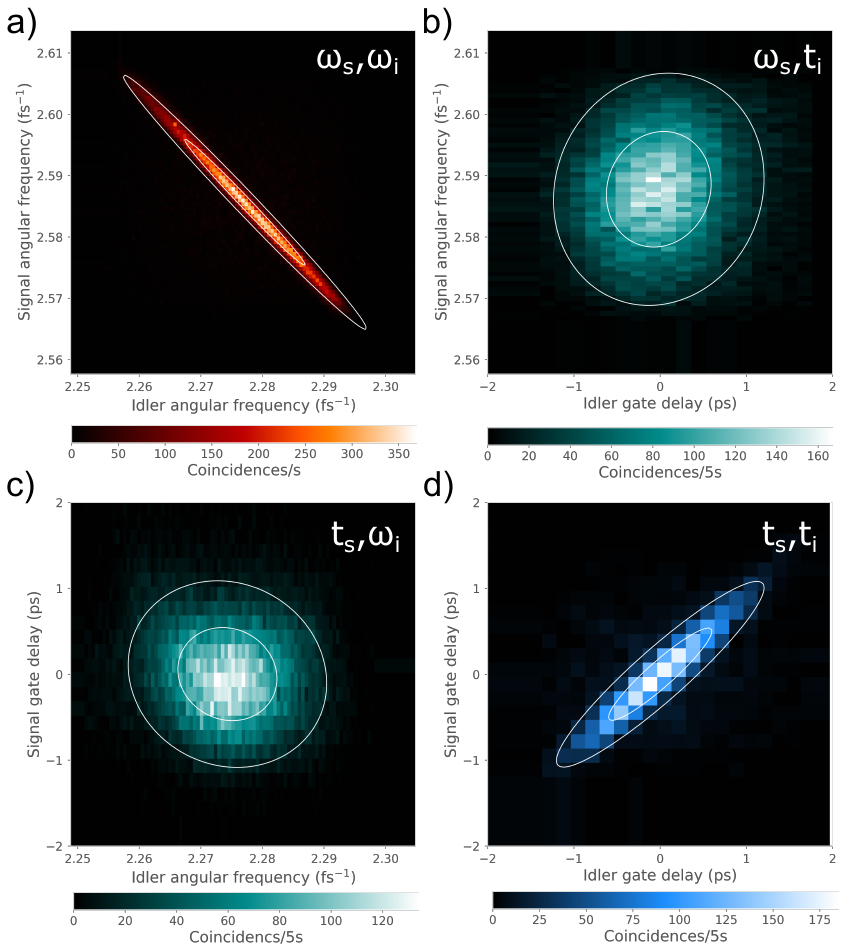} 
     \caption{\footnotesize{ \textbf{Spectral and temporal characterization of ultrafast
     photons.}
     A combination of spectral and temporal measurements are made in
     coincidence in order to measure (a) the joint spectrum, (d) the joint
     temporal intensity, as well as the (b,c) cross-correlations between the
     time (frequency) of the idler and frequency (time) of the signal. (a)
     Frequency anti-correlations with statistical correlation $-0.9951\pm
     0.0001$, are accompanied with (d) positive correlations $0.987\pm0.004$ in
     the signal-idler arrival times.  The time-frequency plots (c,d) show
     little correlations, $(-0.106\pm0.007)$ and ($0.110\pm 0.007$), respectively,
     indicating low dispersion in the signal and idler photons. White lines
     on all plots correspond to $1\sigma$ and $2\sigma$ contours of
     two-dimensional Gaussian fits.}} 
     \label{fig:joint-plots}
   \end{figure}

 Our experimental setup is shown in Fig.~\ref{fig:setup}.  The laser output at
 775~nm is frequency doubled to 387.5~nm in 2~mm of bismuth-borate (BiBO). The
 resulting pump light is spectrally narrowed using a 0.085~nm $(1/\sqrt{e})$
 bandpass filter. Signal and idler photon pairs are created through type-I SPDC
 of the pump in 5~mm of BiBO with central wavelengths of 729~nm and 827~nm,
 respectively. The bandwidths are controlled using tunable spectral edge
 filters after which the photons are coupled to single-mode fibres.  The fibres
 allow for easy switching between spectral measurement, temporal measurement,
 and direct detection. The dispersion of the fibre links is then compensated
 with grating-based pulse compressors.  Spectral measurements are performed
 with grating-based scanning monochromators with a resolution of about 
 0.1~nm.  Temporal measurements are performed through sum-frequency generation
 in 1~mm of type-I BiBO with a strong gate laser pulse with an intensity
 temporal width of 120~fs ($1/\sqrt{e}$), measured using an auto-correlation
 and assuming a Gaussian spectrum.  The upconverted photons are detected after
 passing through spectral bandpass filters which remove the second harmonic
 background of the gate pulse.  We estimate the absolute efficiency of the
 temporal apparatus, including fibre coupling, chirp compensation, and
 upconversion, to be 3\% of the maximum possible.  Detection events for the
 signal and idler are measured in coincidence after they have passed either
 through both spectrometers, both temporal gates, or one of each.  The
 corresponding measured joint spectrum, joint temporal intensity, and
 time-frequency plots, which measure the frequency of one photon in coincidence
 with the arrival time of the other, are shown in Fig.~\ref{fig:joint-plots}.
 Background subtraction has not been employed in the data. 

 For each joint measurement of Fig.~\ref{fig:joint-plots}, the marginal width
 is obtained by fitting the marginals to a one-dimensional Gaussian, while the
 heralded width is obtained taking the average of several slices of the data
 when the frequency or time of one photon is fixed. The statistical
 correlation, $\rho$, is obtained by finding the value that best fits a
 two-dimensional Gaussian with the measured marginals.  Since the finite
 resolution of both spectral and temporal measurements are on the same order of
 magnitude as the spectral and temporal distributions, the measured features
 will be broadened.  To account for this, the fit parameters are deconvolved
 assuming a Gaussian response function~\cite{donohue_spectrally_2016}, and
 these values for the joint spectrum and joint temporal distribution of
 Fig.~\ref{fig:joint-plots}(a,d) are presented in
 Table~\ref{tab:fit-parameters}. 

 The measured joint spectrum shown in Fig.~\ref{fig:joint-plots}(a) exhibits
 strong anti-correlation ($-0.9951\pm0.0001$) in the signal and idler
 frequencies, while the joint temporal intensity of
 Fig.~\ref{fig:joint-plots}(b) shows strong positive correlations
 ($0.987\pm0.004$) in the arrival times of the photons.  We can witness the
 effect of the spectral phase in Fig.~\ref{fig:joint-plots}(b,c), which show
 weak correlations between frequency of one photon and time of arrival of the
 other.  Low correlations in the time-frequency plots may indicate little
 uncompensated dispersion in the experiment (see Supplementary Material). 

  \begin{table}[bt] \centering
    \footnotesize
  \caption{\textbf{Ultrafast two-photon state parameters.}  
  \footnotesize{
  Measured marginals, heralded widths, and correlations of the joint spectrum
  and joint temporal intensity presented Fig.~\ref{fig:joint-plots}(a,d). All
  values are deconvolved to account for the finite resolution of the
  spectrometers and the temporal gate.  Measured properties are 
  widths in standard deviations and error bars are calculated from Monte Carlo
  simulations assuming Poissonian noise.  A more comprehensive list including
  both raw and deconvolved fit parameters can be found in the Supplementary
  Material.}}
  \label{tab:fit-parameters}
    \begin{tabular}{ccc} 
      \hline\hline \multirow{2}{*}{Property} & Joint & Joint temporal\\
      &spectrum& intensity\\
   \hline 
   Signal marginal width 	& $(10.56\pm0.04)~\textrm{ps}^{-1}$  & $(0.537\pm0.009)$~ps  \\ 
   Signal heralded width	& $(1.02\pm0.05)~\textrm{ps}^{-1}$   & $(0.066\pm0.018)$~ps  \\ 
   Idler marginal width 	& $(9.69\pm0.03)~\textrm{ps}^{-1}$   & $(0.587\pm0.015)$~ps  \\
   Idler heralded width		& $(0.94\pm0.04)~\textrm{ps}^{-1}$   & $(0.070\pm0.019)$~ps  \\
   Correlation $\rho$		& $-0.9951\pm0.0001$			 & $0.987\pm0.004$  \\ \hline\hline 
 \end{tabular}
 \end{table}

  \begin{figure}[tb]
     \centering
     \includegraphics[scale=1.0]{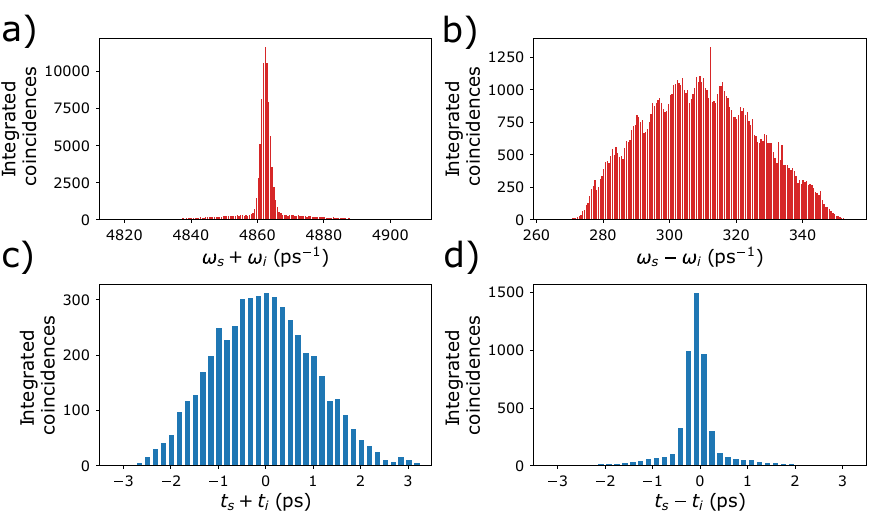} 
     \caption{\footnotesize{\textbf{Histograms of the frequency and time of
     arrival correlations between signal and idler photons.} Coincidences are
     confined to a small region in (a) with
     $\Delta(\omega_s+\omega_i)=(1.429\pm0.006)~\textrm{ps}^{-1}$
     ($1.329\pm0.007~\textrm{ps}^{-1}$ when corrected for the finite resolution
     of the gate) compared to (b) with
     $\Delta(\omega_s-\omega_i)=(18.16\pm0.05)~\textrm{ps}^{-1}$
     ($18.16\pm0.05~\textrm{ps}^{-1}$) indicating strong anti-correlations in
     frequency.  Likewise, coincidences are localized in (d) with
     $\Delta(t_s-t_i)=0.203\pm0.005~\textrm{ps}$ ($0.110\pm0.010$~ps) compared
     to (c) with $\Delta(t_s+t_i)=1.066\pm0.016$~ps ($1.052\pm0.016$~ps)
     corresponding to strong correlations in the time of arrival. From these
     values, we find a joint uncertainty product
     $\Delta(\omega_s+\omega_i)\Delta(t_s-t_i)=0.290\pm0.007 ~(0.15\pm0.01)$.}}
 \label{fig:histograms}
   \end{figure}

 The spectral and timing correlations are further analyzed by binning the data
 presented in Fig.~\ref{fig:joint-plots}(a,d) into histograms based on
 $\omega_1+\omega_2$ and $t_s-t_i$, as well as $\omega_s-\omega_i$ and
 $t_s+t_i$ for comparison, as shown in Fig.~\ref{fig:histograms}. The bin size
 was selected to match the step size of the measurement apparatus.  Gaussian
 fits to the histograms give a joint uncertainty product
 $\Delta(\omega_s+\omega_i)\Delta(t_s-t_i)=(1.429\pm0.006~\textrm{ps}^{-1})
 (0.203\pm0.005~\textrm{ps}) =0.290\pm0.007$, which violates the inequality of
 Eq.~\ref{eq:TBP-inequality} by about 100 standard deviations.  Error bars are
 obtained via Monte-Carlo simulations assuming Poissonian noise.  When
 deconvolved, we find
 $\Delta(\omega_s+\omega_i)\Delta(t_s-t_i)=(1.329\pm0.007~\textrm{ps}^{-1})
 (0.110\pm0.010~\textrm{ps})=0.15\pm0.01$.  The measured uncertainty products
 thus provide a clear witness of energy-time entanglement on ultrafast
 timescales. 

   \begin{figure}[t!]
     \centering
     \includegraphics[scale=1.00]{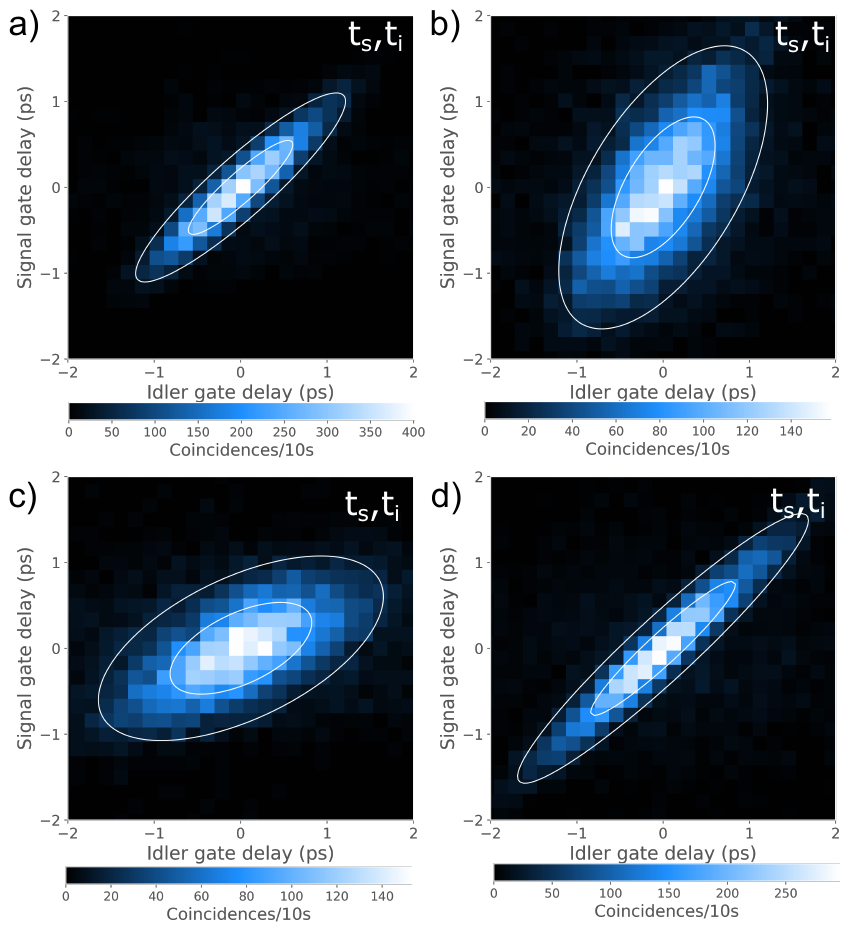} 
     \caption{\footnotesize{ \textbf{Nonlocal dispersion cancellation observed
     in the joint temporal distributions.} Joint temporal intensity for the
     signal and idler pair: (a) without dispersion, (b) with a positive
     dispersion of $A_s=(0.0373\pm0.0015)~\textrm{ps}^2$ on the signal, (c) with a
     negative dispersion of $A_i=(-0.0359\pm0.0014)~\textrm{ps}^2$ on the idler, and
     (d) with both a positive dispersion of $A_s=(0.0373\pm0.0015)~\textrm{ps}^2$ on the
     signal and a negative dispersion of $A_i=(-0.0359\pm0.0014)~\textrm{ps}^2$ on
     the idler.  For each, we measure the uncertainty in the difference in
     arrival times of the signal and idler $\Delta(t_s-t_i)$ and find: (a)
     $0.235\pm0.003$~ps ($0.162\pm0.005$~ps when corrected for the finite resolution of the
     gate), (b) $0.708\pm0.013$~ps ($0.688\pm0.013$~ps), (c) $0.714\pm0.010$~ps
     ($0.693\pm0.011$~ps), (d) $0.245\pm0.004$~ps ($0.175\pm0.006$~ps).  
      We witness nonlocal dispersion cancellation in the timing uncertainty
      $t_s-t_i$ in (d) as the width $\Delta(t_s-t_i)$ remains almost unchanged
      with the one measured in (a).}} \label{fig:dispersion-cancellation}
   \end{figure}

 We now turn to the problem of measuring the impact of dispersion on our
 energy-time entangled state.  We directly observe the effect of applied
 dispersion on the temporal correlations, as presented in the joint temporal
 intensities of Fig.~\ref{fig:dispersion-cancellation}. We control the spectral
 phase of the photons, $\phi(\omega_s,\omega_i)\approx
 A_s(\omega_s-\omega_{s0})^2+A_i(\omega_i-\omega_{i0})^2$, with two grating
 compressors where the chirp parameters $A_s$ and $A_i$ are for the signal and
 idler fields, respectively. We estimate the magnitude of the applied
 dispersion from the geometry of the compressor and the relative position of
 the gratings~\cite{treacy_optical_1969}, and measure the standard deviation
 $\Delta(t_s-t_i)$ of a Gaussian fit from histograms of $t_s-t_i$. 
  
 Starting from the case with no dispersion
 [Fig.~\ref{fig:dispersion-cancellation}(a)], we apply positive dispersion
 $A_s=(0.0373\pm0.0015)~\rm{ps}^2$ to only the
 signal~[Fig.~\ref{fig:dispersion-cancellation}(b)] and negative dispersion
 $A_i=-(0.0359\pm0.0014)~\rm{ps}^2$ to only the
 idler~[Fig.~\ref{fig:dispersion-cancellation}(c)].  In these two cases, we
 observe a large increase in the timing uncertainty $\Delta(t_s-t_i)$ and a
 vertical or horizontal shear of the joint-temporal intensity along the
 corresponding axis.  We then apply the same amount of positive and negative
 dispersion to the signal and idler as
 before~[Fig.~\ref{fig:dispersion-cancellation}(d)], where the dispersion
 applied to the idler is set to minimize the timing uncertainty between the two
 photons.  Here, the timing uncertainty in arrival time $\Delta(t_s-t_i)$ is
 almost unchanged.  This is the signature of nonlocal dispersion cancellation,
 limited by the finite correlations of the initial two-photon state (see
 Supplementary Material).  The temporal marginals in
 Fig.~\ref{fig:dispersion-cancellation}(d) still increase since each side
 remains exposed to a significant amount of dispersion. 
   
 For classical pulses, the effect of dispersion on the correlations in arrival
 times can be expressed as an inequality~\cite{wasak_entanglement-based_2010},
 $\Delta(t_s-t_i)_F^2\ge\Delta(t_s-t_i)_0^2+4A^2/\Delta(t_s-t_i)_0^2$, where
 $\Delta(t_s-t_i)_0$ is the initial difference in detection times, and
 $\Delta(t_s-t_i)_F$ is the final difference with equal and opposite dispersion
 $A$ applied on each side. Under the assumption that the initial state is
 unchirped, taking the measured initial value from
 Fig.~\ref{fig:dispersion-cancellation}(a),
 $\Delta(t_s-t_i)_0=0.235$~ps~(0.162~ps when corrected for the gate resolution), and
 using the average magnitude of the applied dispersion $A=0.0366~\rm{ps}^2$, we
 calculate that the standard deviation in arrival times for classical pulses
 has to be at least $\Delta(t_s-t_i)_F\ge0.390$~ps (0.480~ps). However, the measured
 uncertainty observed in Fig.~\ref{fig:dispersion-cancellation}(d),
 $\Delta(t_s-t_i)=(0.245\pm0.004)$~ps, remains significantly smaller.  The
 experimental apparatus thus provides a direct way to detect this inherently
 quantum effect in a regime inaccessible to current detectors.   
  
 We have directly measured both the temporal and frequency correlations of an
 ultrafast biphoton pulse. Optical gating employed here was critical for
 realizing ultrafast coincidence detection and correspondingly high-resolution
 temporal measurements.  We observe energy-time entanglement via a joint
 time-bandwidth inequality and demonstrate ultrafast nonlocal dispersion
 cancellation of the biphotons with direct and independent detection.  This
 work can be extended to quantum interference measurements on ultrafast
 timescales, and can be combined with temporal imaging to greatly increase the
 versatility of energy-time entangled photons for quantum information
 applications.

 The authors thank M. Mazurek for fruitful discussions and M.  Mastrovich for
 valuable assistance in the laboratory.  This research was supported in part by
 the Natural Sciences and Engineering Research Council of Canada (NSERC),
 Canada Research Chairs, Industry Canada and the Canada Foundation for
 Innovation (CFI).

\bibliography{double_sfg} 
\appendix 
\newpage
\onecolumngrid

\section*{Supplementary Material} 
\section{Additional Experimental Details}

 The experiment uses a titanium-sapphire (Ti:Sapph) laser with an 80~MHz
 repetition rate which produces femtosecond laser pulses 12.5~ns apart centred
 at 775~nm with a $1/\sqrt{e}$ bandwidth of 2.25~nm. These are frequency
 doubled through second harmonic generation in 2~mm  of type-I phasematched
 bismuth borate (BiBO) generating pulsed pump light centred at 387.5~nm with a
 $1/\sqrt{e}$ bandwidth of 0.6~nm and an average power of 900~mW. The resulting
 pump light is spectrally narrowed using a 0.085~nm $(1/\sqrt{e})$ bandpass filter, from
 which we estimate a pump coherence length of approximately 470~fs
 ($1/\sqrt{e}$).  The remaining 300~mW of filtered pump is focussed in 5~mm of
 type-I BiBO for spontaneous parametric downconversion (SPDC). Signal-idler
 photon pairs are created with central wavelengths of 728.6~nm and 827.3~nm,
 respectively, and split with dichroic mirrors. These wavelengths are chosen
 such that the upconverted photon is spectrally far from the laser second-harmonic
 generation (SHG) background. The spectral bandwidths of the photons are
 controlled using a pair of short pass and long pass edge filters on each side.
 Each photon is then coupled into single-mode fibre and can be either be
 spectrally or temporally analyzed.

 Spectral measurements are made with two grating-based scanning
 monochromators (1200 lines/mm), one for each of the two near-infrared (NIR)
 SPDC photons. See Ref.~\cite{donohue_spectrally_2016} for further details.  The
 resolutions of the spectrometers, obtained from the emitted spectra of a Ne-Ar
 calibration lamp, are 0.081~nm and 0.135~nm for the signal and idler,
 respectively, the difference arising from slightly different slit widths in
 each monochromater.  
 
 For the temporal measurements, signal and idler photons are sent through 16.2~m
 and 21.2~m of fibre respectively. Grating-based compressors compensate for
 this chirp and allow variable control over the dispersion.  A polarizing beam
 splitter separates the Ti:Sapph fundamental into two gates pulses.  Due to the
 added propagation in fibre, the signal and idler photons originate
 respectively 7 and 9 pulses behind the gate pulses.  Each photon then
 co-propagates with a gate pulse from the respective side with a spatial
 separation of about 8~mm and is subsequently focussed into 1~mm of type-I
 phasematched BiBO for sum-frequency generation (SFG). The upconverted light,
 with central wavelengths of 375~nm and 400~nm for the signal and idler sides,
 respectively, is recollimated, spectrally filtered with bandpass filters to
 remove second-harmonic background, and then coupled into multimode fibre. The
 SHG background was approximately 10 times higher for the idler SFG compared to
 the signal SFG and therefore, different gate powers were used to maximize the
 signal-to-noise ratios in each arm, with 500~mW for the signal gate pulse and
 200~mW for the idler gate pulse. Both upconverted photons are detected with
 silicon avalanche photodiodes with quantum efficiencies of approximately 30\%
 near 400 nm. The coincidence window for detection events was set to 3~ns.

 The relative separation of the gratings in each compressor is initially
 scanned to cancel the chirp from the fibres.  This is achieved by minimizing
 the upconversion width as a function of the grating separation. The location
 of the minimum defines the centre position of the gratings in the compressor
 where zero dispersion is applied.  The amount of dispersion provided by each
 compressor is then determined from the displacement of the gratings from their
 centre position and their angle with respect to the incident and reflected
 light.  The compressors on the signal and idler arm are thus found to give
 $1315~\rm{fs}^2$ and $1925~\rm{fs}^2$ per mm of displacement, respectively,
 due to the inverse cubic dependence on wavelength~\cite{treacy_optical_1969}.

 Photons were produced at the source at a rate of 673,000 coincidence counts
 per second with $3.4\times10^6$ and $3.5\times10^6$ single-detection events
 per second for the signal and idler, respectively.  The heralded second-order
 coherence of the source, measured with a Hanbury Brown-Twiss
 interferometer~\cite{hanbury_brown_test_1956}, was $g^{(2)}(0)=0.416\pm0.004$
 for the signal and $g^{(2)}(0)=0.415\pm0.003$ for the idler. In general,
 double pair emission will lead to a broad background in the joint spectrum and
 joint temporal intensity. However, due to the tight temporal filtering on both
 sides, we estimate that double pairs contribute to less than 1\% of the
 measured upconverted signal.  After the upconversion on each side,
 approximately 30 coincidence counts ($10,000$ upconverted signal singles and
 $16,000$ upconverted idler singles per second) per second were measured at the
 peak, from which about 0.6 coincidence counts (3,000 and 360 singles) per
 second were background from the second harmonic of the gate pulse. 
 
 See Table~\ref{table:source-parameters} for a list of parameters from the
 joint spectrum, joint temporal intensity, and frequency-time plots in
 Fig.~\ref{fig:joint-plots} of the main text.  See Table
 ~\ref{table:dispersion-parameters} for a collection of parameters for plots of
 the joint temporal intensity in Fig.~\ref{fig:dispersion-cancellation}. Raw
 measurements and deconvolved values are presented in both tables.  The
 deconvolved width $\Delta x$ is defined in terms of the measured width $\Delta
 x_{\rm meas}$ and the resolution of the instrument $\Delta x_{\rm res}$.  For
 example, for the marginals, we have $\Delta x=\sqrt{\Delta x_{\rm meas}^2-\Delta
 x_{\rm res}^2}$.

\begin{table}[h!]
  \centering
\caption{\textbf{Complete fit parameters for joint plots.} Selected properties
of the fits to the joint spectrum, joint temporal intensity, and joint time
frequency plots seen in Fig.~\ref{fig:joint-plots} of the main text. Values in
parentheses are deconvolved from a Gaussian response function.}
  \begin{tabular}{|c|c|c|c|c|c|}
    \hline
 \multicolumn{2}{|c|}{Property}	&  \multirow{2}{*}{Joint-spectrum} & Joint-temporal& Signal frequency& Signal time\\ 
 \multicolumn{2}{|c|}{(Deconvolved)} 	&				&intensity & Idler time & Idler frequency \\ 
 \hline
 \multirow{5}{*}{Signal}& Frequency ($\omega$)	&$2586.9\pm0.4~\rm{ps}^{-1}$		& - 			&	 -			&-  				\\\cline{2-6} 
 	&Marginal 				&$10.57\pm0.04~\rm{ps}^{-1}$ 		& $0.550\pm0.009$~ps 	& $9.43\pm0.05~\rm{ps}^{-1}$   	& $0.533\pm0.003$~ps		\\ 
 	&width					& $(10.56\pm0.04~\rm{ps}^{-1}$) 	& $(0.537\pm0.009$~ps) 	& $(9.42\pm0.05~\rm{ps}^{-1}$) 	& $(0.519\pm0.003$~ps)		\\ \cline{2-6}
 	&Heralded 				& $1.16\pm0.04~\rm{ps}^{-1}$ 		& $0.176\pm0.008$~ps    & $9.4\pm0.2~\rm{ps}^{-1}$	& $0.514\pm0.017$~ps		\\ 
	&width					& $(1.02\pm0.05~\rm{ps}^{-1})$ 		& $(0.066\pm0.018$~ps) 	& $(9.3\pm0.2~\rm{ps}^{-1}$)	& $(0.501\pm0.018$~ps) 		\\\hline 
 \multirow{5}{*}{Idler}& Frequency  ($\omega$)	& $2276.9\pm0.3~ \rm{ps}^{-1}$   	& - 			& - 				& -				\\\cline{2-6} 
 	&Marginal 				& $9.69\pm0.03~ \rm{ps}^{-1}$ 		& $0.600\pm0.015$~ps 	& $0.589\pm0.006$~ps 		& $8.03\pm0.05~\rm{ps}^{-1}$	\\ 
 	&width					& $(9.69\pm0.03~\rm{ps}^{-1}$)		& $(0.587\pm0.015$~ps) 	& $(0.576\pm0.006$~ps)		& $(8.02\pm0.05~\rm{ps}^{-1}$) 	\\\cline{2-6} 
 	&Heralded 				& $1.06\pm0.04~\rm{ps}^{-1}$     	& $0.185\pm0.009$~ps 	& $0.588\pm0.022$~ps		& $7.7\pm0.6~\rm{ps}^{-1}$	\\ 
 	&width					& $(0.94\pm0.04~\rm{ps}^{-1}$)    	& $(0.070\pm0.019$~ps) 	& $(0.576\pm0.022$~ps)		& $(7.7\pm0.6~\rm{ps}^{-1}$)	\\\cline{1-6}
\multicolumn{2}{|c|}{Statistical}	 	& $-0.9939\pm0.0001$			& $0.944\pm0.003$ 	& $0.109\pm0.008$		& $-0.103\pm0.008$	\\ 
\multicolumn{2}{|c|}{Correlation}		& $(-0.9951\pm0.0001)$			& $(0.987\pm0.004)$ 	& $(0.111\pm0.008)$		& $(-0.106\pm0.008)$	\\\hline 
\end{tabular}
\label{table:source-parameters}
\end{table}

\begin{table}[h!]
  \centering
\caption{\textbf{Complete fit parameters for the nonlocal dispersion
cancellation.} Selected properties of the fits to the joint temporal intensity
plots seen in Fig.~\ref{fig:dispersion-cancellation} of the main text. Values
in parentheses are deconvolved from a Gaussian response function.}
  \begin{tabular}{|c|c|c|c|c|c|}
    \hline
    \multicolumn{2}{|c|}{Property} & No dispersion & Positive dispersion &
    Negative dispersion & Opposite \\ 
    \multicolumn{2}{|c|}{(Deconvolved)}& &     on the signal & on the idler& dispersion \\ 
   \hline
   \multirow{4}{*}{Signal} &Marginal	& $0.536\pm0.004 $~ps   & $0.797\pm0.009$~ps    & $0.518\pm0.004$~ps    & $0.764\pm0.007$~ps\\ 
 		&width 			& $(0.523\pm0.004$~ps)  & $(0.788\pm0.009)$~ps  & $(0.504\pm0.004)$~ps  & $(0.754\pm0.007)$~ps\\ \cline{2-6}
    		&Heralded 		& $0.206\pm0.010 $~ps   & $0.619\pm0.040$~ps    & $0.405\pm0.020$~ps   	& $0.230\pm0.030$~ps\\ 
 		&width 			& $(0.132\pm0.012$~ps)  & $(0.599\pm0.040)$~ps  & $(0.384\pm0.021)$~ps 	& $(0.160\pm0.030)$~ps\\ \hline
    \multirow{4}{*}{Idler} &Marginal    & $0.592\pm0.006$~ps    & $0.585\pm0.005$~ps    & $0.795\pm0.010$~ps    & $0.820\pm0.010$~ps\\ 
 		&width			& $(0.580\pm0.006)$~ps  & $(0.572\pm0.005)$~ps 	& $(0.786\pm0.010)$~ps  & $(0.811\pm0.010)$~ps\\\cline{2-6} 
     		&Heralded		& $0.223\pm0.012$~ps    & $0.448\pm0.013$~ps 	& $0.613\pm0.004$~ps   	& $0.231\pm0.015$~ps\\ 
 		&width			& $(0.143\pm0.014)$~ps  & $(0.428\pm0.014)$~ps  & $(0.592\pm0.004)$~ps 	& $(0.160\pm0.020)$~ps\\\hline 
 \multicolumn{2}{|c|}{Statistical}	& $0.912\pm0.003$     	& $0.589\pm0.009$ 	& $0.56\pm0.01$   & $0.951\pm0.002$\\ 
 \multicolumn{2}{|c|}{Correlation}	& $(0.956\pm0.003)$   	& $(0.609\pm0.009)$ 	& $(0.58\pm0.01)$ & $(0.973\pm0.002)$\\\hline 
\end{tabular}
\label{table:dispersion-parameters}
\end{table}
\clearpage

\section{Signatures of energy-time entanglement}
 In this section, we calculate the joint-uncertainty product
 $\Delta(\omega_s+\omega_i)\Delta(t_s-t_i)$ using a model for a two-photon
 state with variable energy-time
 entanglement~\cite{prevedel_classical_2011,donohue_spectrally_2016}.  We show
 that entangled quantum states can violate the inequality of
 Eq.~\ref{eq:TBP-inequality} and describe the time-bandwidth products (TBP) of
 this state.  This requires calculating both the joint spectral intensity and
 the joint temporal intensity.

 Consider the correlated two-mode state,  
 \begin{align}
     \ket{\psi}=\int
     d\omega_sd\omega_iF(\omega_s,\omega_i)
     a^{\dagger}_{\omega_s}a^{\dagger}_{\omega_i}\ket{0},
     \label{eq:SPDC_state}
   \end{align}
 with the normalized joint-spectral amplitude expressed in Gaussian form as, 
 \begin{align}
   F\left( \omega_s,\omega_i \right) = 
   \frac{1}{\sqrt{2\pi\sigma_{\omega_s}\sigma_{\omega_i} } \left( 1 - \rho_{\omega}^2 \right)^{1/4}} 
   \exp \left( { -\frac{1}{{2\left( {1 - {\rho_{\omega} ^2}} \right)}}\left[ {\frac{{{{\left(
     {{\omega _s} - {\omega _{s0}}} \right)}^2}}}{{2{\rm{\sigma }}_{\omega_s}^2}} 
     + \frac{{{{\left( {{\omega _i} - {\omega _{i0}}} \right)}^2}}}{{2\sigma
       _{\omega_i}^2}} 
 - \frac{\rho_{\omega} \left(\omega_s-\omega_{s0}\right)\left(\omega _i-\omega
 _{i0}\right)}{\sigma_{\omega_s}\sigma_{\omega_i}}} \right]}
 \right).
 \label{eq:JSA}
  \end{align}
  In the two-mode state of Eq.~\ref{eq:JSA}, there are two relevant length
  scales for the signal and idler, which we refer to as the marginal width,
  $\Delta\omega^{(m)}$ , and the heralded or coincident width,
  $\Delta\omega^{(h)}$, and where, here, $\Delta
  x=\sqrt{\left<x^2\right>-\left<x\right>^2}$ refers to the intensity standard
  deviation or $1/\sqrt{e}$ width of the variable $x$.  The marginal widths in
  the equation are obtained by taking the marginal over one photon and tracing
  out or ignoring the other, while the heralded widths are obtained by fixing
  the frequency of either the signal or idler to its central frequency
  ($\omega_s\rightarrow\omega_{s0}$ or $\omega_i\rightarrow\omega_{i0}$). Using
  Eq.~\ref{eq:JSA} above, we find,
 \begin{align}
    \label{eq:Dwm}
   &\Delta\omega^{(m)}_{s,i}=\sigma_{\omega_{s,i}}\\
    \label{eq:Dwh}
   &\Delta\omega^{(h)}_{s,i}=\sqrt{1-\rho_{\omega}^2}\sigma_{\omega_{s,i}}
 \end{align}
  The correlation parameter $\rho_{\omega}=\Delta(\omega_s\omega_i)/\Delta \omega_s
  \Delta \omega_i$ describes the statistical correlations between the frequency
  of the signal and idler modes and is related to the purity of the partial
  trace, $P=\sqrt{1-\rho_{\omega}^2}$. When $\rho_{\omega}=0$, the joint-spectral amplitude
  $F(\omega_s,\omega_i)$ factorizes and the state is separable, whereas when
  $\rho_{\omega}\rightarrow-1$, the photons are perfectly anti-correlated in frequency
  and when $\rho_{\omega}\rightarrow\-1$, they are perfectly correlated.

 The joint temporal amplitude is obtained by taking the Fourier transform
 of the joint spectral amplitude,
 \begin{align}
   \begin{split}
   f\left( {{t_s},{t_i}} \right) &=
   \int d{\omega _i}d{\omega _s}F\left( {{{\rm{\omega }}_i},{{\rm{\omega }}_s}}
   \right){e^{i{\omega _i}{t_i}}}{e^{i{\omega _s}{t_s}}}\\
   &=\frac{1}{\pi}\sqrt {2\pi {{\rm{\sigma }}_{\omega_s}}{\sigma
     _{\omega_i}}}(1-\rho_{\omega}^2)^{1/4} \exp\left(- t_s^2\sigma_{\omega_s}^2  - t_i^2\sigma_{\omega_i}^2
     - 2t_st_i\rho_{\omega} \sigma_{\omega_s}\sigma_{\omega_i}
 -i\left(t_s\omega_{s0} + t_i\omega_{i0}\right)\right).
   \end{split}
   \label{eq:JTA}
  \end{align}
  Equation~\ref{eq:JTA} can be recast as a two-dimensional Gaussian in the form of
  Eq.~\ref{eq:JSA} and in doing so, we obtain expressions for the
  marginal pulse width $\Delta t^{(m)}$ and the heralded pulse width $\Delta
  t^{(h)}$ for the signal and idler, as well as the statistical correlations
  $\rho_t$ between the time of arrival of the photons,
  \begin{align}
    \label{eq:Dtm}
    \Delta{t^{(m)}_{s,i}}&=\frac{1}{2\sqrt{1-\rho_{\omega}^2}\sigma_{\omega_{s,i}}}\\
    \label{eq:Dth}
    \Delta{t^{(h)}_{s,i}}&=\frac{1}{2\sigma_{\omega_{s,i}}}\\
  \rho_t&=-\rho_{\omega}.
  \end{align}
  We observe that the marginal pulse width $\Delta t^{(m)}$ is inversely
  proportional to the heralded bandwidth
  $\sqrt{1-\rho_{\omega}^2}\sigma_{\omega}$ and heralded pulse widths $\Delta
  t^{(h)}$ is inversely proportional to the marginal bandwidth
  $\sigma_{\omega}$. In addition, the statistical correlations in the temporal
  intensity, $\rho_t$, are reversed from those in the spectral intensity,
  $\rho_\omega$. 

  \subsubsection{Joint-uncertainty product}

  Using both joint amplitude functions of Eq.~\ref{eq:JSA} and
  Eq.~\ref{eq:JTA}, we can calculate the variance in the sum of the frequencies
  of the signal and idler,  
  \begin{align}
    \begin{split}
  {\rm{\Delta }}{\left( {{\omega_s} + {\omega_i}} \right)^2}
       &=\sigma_{\omega_s}^2 +
       2\rho_{\omega} {\sigma_{\omega_i}}{\sigma_{\omega _s}} + \sigma_{\omega _i}^2,
     \end{split}
   \end{align}
   and variance in the difference in time of arrival,
  \begin{align}
    \label{eq:dt2t1FL}
  &{\rm{\Delta }}{\left( {{t_s} - {t_i}} \right)^2}
   = \frac{{\sigma_{\omega _s}^2 + 2\rho_{\omega} {\sigma_{\omega _i}}{\sigma_{\omega
   _s}} + \sigma_{\omega _i}^2}}{{4\left( {1 - {\rho_{\omega} ^2}} \right)\sigma_{\omega
   _i}^2\sigma_{\omega_s}^2}},
  \end{align}
  in order to obtain the joint uncertainty product,
  \begin{align}
  {\rm{\Delta }}{\left( {{\omega_s} + {\omega_i}} \right)} {\rm{\Delta
  }}{\left( {{t_s} - {t_i}} \right)}=
    \sqrt{\frac{\left({\sigma_{\omega _s}^2 + 2\rho_{\omega} {\sigma_{\omega _i}}{\sigma_{\omega
   _s}} + \sigma_{\omega _i}^2}\right)^2}{{4\left( {1 - {\rho_{\omega} ^2}} \right)\sigma_{\omega
   _i}^2\sigma_{\omega_s}^2}} }.
    \label{}
  \end{align}
  If the bandwidths $\sigma_{\omega_s}=\sigma_{\omega_i}$ are equal, then for the state above, the
 joint uncertainty product is
 $\Delta(\omega_s+\omega_i)\Delta(t_s-t_i)=\sqrt{(1+\rho_{\omega})/(1-\rho_{\omega})}$. In this
 case, the joint uncertainty product for the transform limited two-photon state
 depends entirely on the frequency correlation parameter $\rho_{\omega}$.   When
 $\rho_{\omega}<0$, the state clearly violates Eq.~\ref{eq:TBP-inequality}; the
 simultaneous correlations in frequency and time are stronger than those
 achievable with classical pulses and the state is energy-time entangled. When
 $\rho_{\omega}=0$, the state satisfies the equality as it is separable.  The presence
 of dispersion on either photon increases the overall product. Spectral phase
 stretches the temporal profile of the photons and increases the uncertainty
 $\Delta(t_s-t_i)$ in their arrival time without affecting the uncertainty in
 the bandwidth $\Delta(\omega_s+\omega_i)$.  If the photons are positively
 correlated $\rho_{\omega}>0$, a different joint-uncertainty product can be used to
 verify entanglement, namely $\Delta(\omega_s-\omega_i)\Delta(t_s+t_i)$ which
 is also always greater than or equal to one for separable states.

 \subsubsection{Time-bandwidth products}
 From the joint spectral and joint temporal amplitude functions, we can also obtain
 a set of time-bandwidth products (TBP) for the individual modes. For a
 classical pulse, the TBP must satisfy the uncertainty relation, 
 \begin{align}
 \Delta \omega \Delta t\geq1/2.
    \label{}
 \end{align}
 On the other hand, for correlated photons, there are four possible
 time-bandwidth products. The first two time-bandwidth products of the
 individual photons compare the marginal (heralded) bandwidth to the heralded
 (marginal) temporal pulse width. For the Fourier limited two-photon state
 presented above, using Eqs.~\ref{eq:Dwm}, \ref{eq:Dwh}, \ref{eq:Dtm}, and
 \ref{eq:Dth}, they are, 
 \begin{align}
    \label{eq:TBP} 
      \Delta \omega^{(m)}\Delta t^{(h)}=1/2\\
      \Delta \omega^{(h)}\Delta t^{(m)}=1/2. 
 \end{align}
 These TBPs take place of the classical time-bandwidth products, and hold
 regardless of the amount of entanglement in the system. In the presence of a
 nonzero spectral phase, the temporal widths will increase whereas the
 frequency widths will remain the same, and the TBP will only get larger. The
 value 1/2 is thus a minimum which is attained when there is no spectral phase. 

 The last two TBPs compare both marginal widths and both heralded widths, and
 we find, 
  \begin{align}
    \Delta \omega^{(m)}\Delta t^{(m)}=\frac{1}{2}\frac{1}{\sqrt{1-\rho_{\omega}^2}}\\
    \Delta \omega^{(h)}\Delta t^{(h)}=\frac{1}{2}\sqrt{1-\rho_{\omega}^2}. 
  \end{align}
 These TBPs depend on the strength of the frequency correlations
 $\rho_{\omega}$. Both reduce to 1/2 when there are no correlation and the
 state is spectrally pure, $\rho_{\omega}=0$.  The marginal TBP,
 $\Delta\omega^{(m)}\Delta t^{(m)}$, will increase for a correlated state
 $0<|\rho_{\omega}|<1$, whereas the heralded TBP, $\Delta\omega^{(h)}\Delta
 t^{(h)}$, will decrease. Energy-time entangled states can have a heralded TBP
 much smaller than 1/2 when $\rho_{\omega}<0$. Since this is forbidden for
 classical pulses, it can also be used as a measure of entanglement, and has
 been shown to be directly related to the spectral purity of the
 state~\cite{brecht_characterizing_2013}.  Similarly to the two previous TBPs,
 both the marginal TBP and the heralded TBP will increase in the presence of
 nonzero spectral phase. 
 
 \section{Energy-time entanglement with dispersion}
 We next analyze the effect of dispersion on the energy-time entangled state in
 order to determine its effect on the joint-uncertainty product and the
 conditions under which nonlocal dispersion cancellation can be observed.
 Starting with the joint spectral amplitude in Eq.~\ref{eq:JSA}, we apply
 dispersion to both photons,
 \begin{align}
   F(\omega_s,\omega_i)\rightarrow
   F(\omega_s,\omega_i)e^{i\phi(\omega_s,\omega_i)},
   \label{}
 \end{align}
 and assume spectral phase has the separable form,
 $\phi(\omega_s,\omega_i)=A_i(\omega_i-\omega_{i0})^2+A_s(\omega_s-\omega_{s0})^2$,
 with chirp parameters $A_i$ and $A_s$.  The presence of spectral phase 
 will not affect any of the spectral intensity measurements. It will,
 however, stretch the temporal marginal of the photons, and we can witness this
 change in the increase of the marginal pulse widths, 
\begin{align}
  \Delta{t_{s,i}}^{(m)}&=\sqrt{\frac{1}{{4(1 - {\rho_{\omega} ^2})\sigma_{s,i}^2}} +
  4{A_{s,i}}^2{\sigma _{\omega_{s,i}}}^2}.
  \label{eq:dtm}
 \end{align}
 The marginal width in time with dispersion has two terms. The first term is
 the Fourier limited marginal width found in Eq.~\ref{eq:Dtm}. When the chirp
 parameter is nonzero, $A_{s,i}\neq0$, for the signal or the idler, we see an
 increase in the corresponding marginal due to the second term
 $4A_{s,i}^2\sigma_{s,i}^2$, regardless of the sign of $A_{s,i}$. Moreover, the
 dispersion $A$ is applied to the entire marginal frequency bandwidth
 $\sigma_{\omega}$.  On the other hand, the joint temporal properties of the
 photons do depend on the relative sign of $A_i$ and $A_s$, and we can observe
 this in the heralded pulse width $\Delta{t_{s}}^{(h)}$ or in the variance of
 the difference in time of arrival of the signal and idler ${\rm{\Delta
 }}{\left( {{t_s} - {t_i}} \right)^2}$.  For example, the signal heralded pulse
 width under dispersion is, 
 \begin{align}
   \label{eq:hpulsewidth}
   \Delta{t_{s}}^{(h)}&=\sqrt{\frac{1}{{4\sigma_{s}^2}} +
   4A_{s}^2(1-\rho_{\omega}^2)\sigma_{s}^2+
   \frac{4\rho_{\omega}^2\left(A_s\sigma_s^2+A_i\sigma_i^2\right)^2}{\sigma_s^2(1+16A_i^2(1-\rho^2)\sigma_i^4)}},
  \end{align}
 the idler heralded pulse width $\Delta t_i^{(h)}$ is obtained from
 Eq.~\ref{eq:hpulsewidth} by exchanging all subscripts $s$ with subscripts $i$,
 and the variance ${\rm{\Delta }}{\left( {{t_s} - {t_i}} \right)^2}$ is, 
 \begin{align}
  {\rm{\Delta }}{\left( {{t_s} - {t_i}} \right)^2}
   &= \frac{{\sigma _s^2 + 2\rho_{\omega} {\sigma _i}{\sigma _s} + \sigma
 _i^2}}{{4\left( {1 - {\rho_{\omega} ^2}} \right)\sigma _i^2\sigma _s^2}} + 4{\left(
   {{A_s}{\sigma _s} + {A_i}{\sigma _i}} \right)^2} - 8{A_i}{A_s}\left( {1 +
   \rho_{\omega} } \right){\sigma _i}{\sigma _s}.
    \label{eq:d(ts-ti)}
  \end{align}
 We focus on the variance, $\Delta(t_s-t_i)^2$, in Eq.~\ref{eq:d(ts-ti)} as the
 other two heralded pulse widths have a similar structure. We find that it
 consists of three distinct terms: the first is the Fourier-limited variance
 when no chirp is applied as in Eq.~\ref{eq:dt2t1FL}, the second is the origin
 of the nonlocal dispersion cancellation as it goes to 0 when
 $A_s\sigma_s=-A_i\sigma_i$, and the third results from the finite correlations
 in the model and also goes to zero for perfect anti-correlations
 $\rho_{\omega}\rightarrow-1$. The variance in Eq.~\ref{eq:d(ts-ti)} can only
 increase in the presence of dispersion, and therefore, the same holds for the
 joint-uncertainty product of Eq.~\ref{eq:TBP-inequality}.   
  
 In order to observe complete nonlocal dispersion cancellation for frequency
 anti-correlated photons, two conditions must be met: the dispersion must be
 opposite in sign with ratios given by $A_s\sigma_s=-A_i\sigma_i$, and the
 photons must be perfectly anti-correlated in frequency, $\rho_{\omega}=-1$. In
 the present experiment,  the first condition is satisfied by setting the
 signal chirp $A_s$ and finding the idler chirp $A_i$ that minimizes the
 uncertainty in arrival time.  However, since we apply dispersion to photons
 with finite correlations, $\rho_{\omega}>-1$, the second condition isn't met
 exactly, and this contributes to increasing the spread in arrival times
 $\Delta(t_s-t_i)$ as observed in the imperfect cancellation of
 Fig.~\ref{fig:dispersion-cancellation}(d). 

 \section{Measurements of frequency-time correlations as an indication of dispersion}

 We now illustrate how the measurements of the cross-correlation in the
 frequency of one photon and time of arrival of the other provide information
 on the dispersion.  A temporal measurement is applied to the signal photon
 $\omega_s$ and a spectral measurement is applied to the idler photon
 $\omega_i$. The temporal measurement is modelled as a convolution of the input
 signal photon spectra with the gate pulse,
 \begin{align}
 {G}\left( {{\omega _g},{\tau}} \right) 
 =\frac{1}{{{{\left( {2\pi {\rm{\sigma }}_g^2} \right)}^{\frac{1}{4}}}\;}}\;
 \exp \left( { - \frac{{{{\left( {{\omega _g} - {\omega _{g0}}}
 \right)}^2}}}{{4{\rm{\sigma }}_g^2}}} +i\tau\left( \omega_g -
 \omega_{g_0}\right)\right)
 \end{align}
 which has centre frequency $\omega_g$, marginal bandwidth $\sigma_g$, and
 delay $\tau$. The upconverted photon at frequency
 $\omega_3=\omega_s+\omega_g$, is then measured in coincidence with the
 spectrally filtered idler photon $\omega_2$, and the probability of measuring
 a coincidence is, 
 \begin{align}
  S\left( {{\tau _s},{\omega _i}} \right) 
 = \int d{\omega _3}{\left| {\smallint d{\omega _s}G\left( {{\omega _3} - {\omega _s},{\tau _s}} \right)
 {\rm{\Phi_{\rm{SFG}} }}\left( {{\omega _s},{\omega _3} - {\omega _s},{\omega
 _3}} \right)\;F\left( {{\omega _s},{\omega _i}} \right)} \right|^2},
   \label{eq:spectrogram}
 \end{align}
 where  $\Phi_{\rm{SFG}}$ is the phasematching function of the sum-frequency
 generation process in the temporal measurement and $F(\omega_s,\omega_i)$ is
 the joint spectral amplitude.  For simplicity, we assume the
 phasematching is infinitely broad $\Phi_{\rm{SFG}}\approx1$.  While this
 assumption isn't strictly valid, the phasematching for the crystals used in
 the experiment isn't strong enough to change the intuition presented here. 

 Since the convolution of two Gaussians is a Gaussian, we can re-express
 $S(\tau_s,\omega_i)$ of Eq.~\ref{eq:spectrogram} as a
 two-dimensional Gaussian such as in Eq.\ref{eq:JSA}, with the  marginal
 bandwidth $\Delta{\omega_i}^{(m)}$, marginal pulse width $\Delta t_s^{(m)}$,
 and statistical correlation $\rho_f$ as follows,
\begin{align}
  \label{eq:mbandwidth}
  \Delta{\omega_i}^{(m)}&=\sigma_s\\
  \label{eq:mpulsewidth}
  \Delta{t_s}^{(m)}&=\sqrt{\frac{1}{{4\sigma_g^2}} + \frac{1}{{4(1 - {\rho_{\omega} ^2})\sigma _s^2}} +
 4{A_s}^2{\sigma _s}^2}\\
 \rho_f&=\frac{{-4{A_s}\rho_{\omega} \sqrt {1 - {\rho_{\omega} ^2}} {\sigma _g}{\sigma
  _s}^2}}{ {\sqrt { {\left( {1 - {\rho_{\omega} ^2}} \right){\sigma _s}^2 +
  {\sigma _g}^2\left( {1 + 16{A_s^2}\left( {1 - {\rho_{\omega} ^2}} \right){\sigma
  _s}^4} \right)} } }}.
  \label{eq:rhof}
\end{align}
 We see that the marginal bandwidth of the idler $\Delta{\omega_i}^{(m)}$ in
 Eq.~\ref{eq:mbandwidth} is independent to the chirp $A_i$ as the spectral
 measurement is independent of phase. In the limit of zero chirp on the signal,
 $A_s=0$, the marginal pulse width $\Delta{t_s}^{(m)}$ in
 Eq.~\ref{eq:mpulsewidth} is a quadrature sum of the gate pulse width
 $1/2\sigma_g$ and the coherence length of the signal
 $1/(2\sqrt{1-\rho_{\omega}^2}\sigma_s)$ from Eq.~\ref{eq:Dtm}, and the frequency of the
 idler and time of arrival of the signal are uncorrelated, $\rho_f=0$ in
 Eq.~\ref{eq:rhof}. When $A_s\neq0$, the signal marginal is stretched by the
 presence of the extra term $4A_s^2\sigma_s^4$ in Eq.~\ref{eq:mpulsewidth}, the
 same term that appears in Eq.~\ref{eq:dtm}, and the correlations increase with
 $-A_s$ in Eq.~\ref{eq:rhof}. 

 The effect of phase-matching is now briefly considered.  Second-order
 phasematching effects describes photons of different frequencies walking off
 from each other inside the crystal.  In the SFG process considered here, a
 photon and gate pulse in the NIR are upconverted to produce a higher energy
 photon in the ultraviolet.  When the upconverted photon walks off from the
 signal and gate, the upconversion becomes partially mode selective and is no
 longer sensitve to all frequencies\cite{brecht_photon_2015}.  Since the
 photons are correlated in frequency, the effective frequency filtering on one
 side has the effect of reducing the measured spectral bandwidth of the photon
 on the other side. We observe this slight reduction when comparing the
 marginal bandwidth in the joint spectrum and frequency-time plots in
 Table~\ref{table:source-parameters}. 

\section{Additional Experimental Results}
\subsubsection{Time-bandwidth products}
 The four additional measured TBPs are presented in
 Table~\ref{table:tb-products}.  We find that the first two TBPs approach the
 value of 1/2 obtained for a Fourier limited two-dimensional Gaussian pulse.
 The difference between the measured values and the value of 1/2 could be due
 to a few reasons.  Uncompensated dispersion will increase both TBPs. The
 time-frequency plots do exhibit small correlations in
 Fig.~\ref{fig:joint-plots}, which would also arise from a nonzero spectral
 phase.  In addition, bandwidth filtering will increase the heralded width
 $\Delta t^{(h)}$ as it depends directly on the marginal spectrum and thus
 further increase $\Delta\omega^{(m)}\Delta t^{(h)}$.  We observe a small
 amount of spectral clipping from the edge filters in the source which will
 reduce the spectral bandwidth of the photons.  Any bandwidth filtering from
 the grating compressors would have the same effect.  Moreover, since the
 heralded width depends on the measurements on both the signal and idler side,
 errors associated with it tend to be larger.  This error translates to the
 deconvolved value, and the measured error on the TBPs involving $\Delta
 t^{(h)}$.   

 When observing the other two TBPs, we find that the marginal TBP
 $\Delta\omega^{(m)}\Delta t^{(m)}$ is much larger than the minimum of 1/2.
 This is consistent with either a mixed state or a spectrally correlated state.
 The heralded TBP $\Delta\omega^{(h)}\Delta t^{(h)}$ is smaller than the
 classically allowed value of 1/2, providing yet another confirmation that the
 photons exhibit energy-time entanglement.

 \begin{table}[b!]
     \centering
     \caption{\textbf{Time-bandwidth products.} The four time-bandwidth
     products of the signal and idler photons from SPDC are obtained from the
     marginal and heralded widths of Fig.~\ref{fig:joint-plots}(a,d).  Values
   in parentheses are deconvolved with a Gaussian response function.}
     \begin{tabular}{ccc}
       \hline\hline
    TBP & Signal & Idler \\ 
    (Deconvolved) & &\\
    \hline
   \multirow{2}{*}{$\Delta\omega^{(m)}\Delta t^{(h)}$}	&$1.64\pm0.07$		& $1.64\pm0.09$ \\ 
							&$(0.62\pm0.15)$	& $(0.62\pm0.16)$\\\hline
    \multirow{2}{*}{$\Delta\omega^{(h)}\Delta t^{(m)}$}	&$0.63\pm0.03$  	& $0.64\pm0.03$\\
							&$(0.55\pm0.03)$  	& $(0.55\pm0.03)$\\\hline
    \multirow{2}{*}{$\Delta\omega^{(m)}\Delta t^{(m)}$}	&$5.16\pm0.07$ 		& $5.3\pm0.1$ \\
    							&$(5.03\pm0.07)$ 	& $(5.2\pm0.1)$\\\hline
    \multirow{2}{*}{$\Delta\omega^{(h)}\Delta t^{(h)}$}	&$0.20\pm 0.01$ 	& $0.20\pm0.01$\\
							&$(0.07\pm 0.02)$ 	& $(0.07\pm0.02)$\\
\hline\hline
   \end{tabular}
   \label{table:tb-products}
   \end{table}

\end{document}